\documentclass[lettersize,journal]{IEEEtran}
\usepackage{amsmath,amsfonts}
\usepackage{array}
\usepackage{textcomp}
\usepackage{stfloats}
\usepackage{verbatim}
\usepackage{cite}
\usepackage{amsmath}
\usepackage{comment}
\usepackage{multirow}
\usepackage[ruled]{algorithm2e}
\usepackage{graphicx}
\usepackage{lipsum}
\usepackage{url}

\usepackage{subfigure}
\usepackage{caption}
\usepackage{booktabs,chemformula}
\usepackage{rotating}
\usepackage{fancyhdr}
\usepackage{gensymb}
\usepackage{enumitem}
\usepackage{color}
\usepackage{xcolor,colortbl}

\usepackage{algpseudocode}
\usepackage{algcompatible}
\definecolor{Gray}{gray}{0.85}
\usepackage{blindtext} 
\usepackage{subfiles}
\usepackage[T1]{fontenc}
\usepackage[utf8]{inputenc}
\begin{document}

\title{Characterization and Optimization of Integrated Silicon-Photonic Neural Networks under Fabrication-Process Variations}

\author{Asif Mirza$^*$,~\IEEEmembership{Student Member,~IEEE}, Amin Shafiee$^*$, Sanmitra Banerjee$^*$,~\IEEEmembership{Student Member,~IEEE}, Krishnendu Chakrabarty,~\IEEEmembership{Fellow,~IEEE}, Sudeep Pasricha,~\IEEEmembership{Senior Member,~IEEE}, and Mahdi Nikdast,~\IEEEmembership{Senior Member,~IEEE} 
\thanks{$^*$Authors with equal contribution. }
\thanks{This work was supported in part by the National Science Foundation (NSF) under grant numbers CCF-1813370, CCF-2006788, and CNS-2046226.}
\thanks{A. Mirza, A. Shafiee, S. Pasricha, and M. Nikdast are with the Department of Electrical and Computer Engineering at Colorado State University, Fort Collins, CO.} \thanks{S. Banerjee and K. Chakrabarty are with the Department of Electrical and Computer Engineering at Duke University, Durham, NC.}}



\maketitle

\begin{abstract}

Silicon-photonic neural networks (SPNNs) have emerged as promising successors to electronic artificial intelligence (AI) accelerators by offering orders of magnitude lower latency and higher energy efficiency. Nevertheless, the underlying silicon photonic devices in SPNNs are sensitive to inevitable fabrication-process variations (FPVs) stemming from optical lithography imperfections. Consequently, the inferencing accuracy in an SPNN can be highly impacted by FPVs---e.g., can drop to below 10\%--- the impact of which is yet to be fully studied. In this paper, we, for the first time, model and explore the impact of FPVs in the waveguide width and silicon-on-insulator (SOI) thickness in coherent SPNNs that use Mach--Zehnder Interferometers (MZIs). Leveraging such models, we propose a novel variation-aware, design-time optimization solution to improve MZI tolerance to different FPVs in SPNNs. Simulation results for two example SPNNs of different scales under realistic and correlated FPVs indicate that the optimized MZIs can improve the inferencing accuracy by up to 93.95\% for the MNIST handwritten digit dataset---considered as an example in this paper---which corresponds to a $<$0.5\% accuracy loss compared to the variation-free case. The proposed one-time optimization method imposes low area overhead, and hence is applicable even to resource-constrained designs.  

\end{abstract}


\section{Introduction} 
\label{sec:: 1-Introduction}
\IEEEPARstart{M}{achine} learning algorithms are being utilized in a wide range of applications ranging from autonomous driving, real-time speech translation, and network anomaly detection to pandemic growth and trend prediction. With the rising demand for advanced neural networks to address even more complex problems, artificial intelligence (AI) accelerators need to consistently deliver better performance and improved accuracy while being energy-efficient. Unfortunately, in the post-Moore's law era, electronic AI accelerator architectures face fundamental limits in their processing capabilities due to the limited bandwidth and low energy efficiency of metallic conductors \cite{pasricha2020survey}. Silicon photonics can alleviate these bottlenecks by enabling chip-scale optical interconnects with ultra-high bandwidth, low-latency, and energy-efficient communication \cite{pasricha2020survey}, and light-speed chip-scale optical computation \cite{AI_survey}. Leveraging silicon-photonic-enabled optical interconnect and computation, many integrated silicon-photonic neural networks (SPNNs) have been recently proposed \cite{AI_survey}.



Prior work has shown that the complexity of matrix-vector multiplication can be reduced from $O(N^2)$ to $O(1)$ in SPNNs \cite{cheng2020silicon}. Also, as SPNNs use photons for computation, there is negligible latency associated with inferencing. Such benefits have positioned SPNNs as a promising alternative to traditional electronic neural networks \cite{cartlidge2020optical}. 
Nevertheless, fabrication-process variations (FPVs) due to inevitable optical lithography imperfections lead to undesired changes in the critical dimensions of SPNNs' building blocks (e.g., waveguide width and thickness in Mach--Zehnder interferometers (MZIs) in coherent SPNNs), imposing inaccuracies during matrix-vector multiplication. In \cite{banerjee2021modeling}, we have shown that random uncertainties due to FPVs and thermal crosstalk can result in up to a catastrophic 70\% accuracy loss, even in mature fabrication processes. Existing approaches for improving the resilience of SPNNs against FPVs largely rely on compensating their impact by individually calibrating MZIs in the network \cite{zhu2020countering}. However, such solutions impose additional calibration (i.e., tuning) power overhead and are not scalable as their complexity rapidly grows with the number of MZIs in SPNNs. 


The key contribution of this paper is in developing, to the best of our knowledge, the first comprehensive analysis of the impact of physical-level FPVs on coherent SPNN performance. We consider variations in the waveguide width and silicon-on-insulator (SOI) thickness, and based on realistic FPV maps developed using different correlation lengths in the variations and experimental measurements. We model the impact of FPVs at the device level for MZI performance and at the network level for arrays of cascaded MZIs in SPNNs. At the system level, we explore the impact of variations on SPNN inferencing accuracy while considering different FPVs and variation correlation lengths. Leveraging our detailed device-level models, we also develop a novel design optimization solution to improve MZI performance in SPNNs under FPVs. Our simulation results for two example SPNNs (with 1380 and 20,580 phase shifters) under FPVs---considered using realistic and correlated wafer variation maps---show that while inferencing accuracy can drop to as low as 6.68\% under waveguide width and SOI thickness variations, using our optimized MZIs can improve the accuracy to 93.95\%. This is within 0.5\% of the nominal inferencing accuracy (in the absense of FPVs) of the example SPNNs considered.    



The rest of the paper is organised as follows. Section~\ref{sec:: 2-Background} presents a background on MZIs and SPNNs and a summary of prior related work. In Section~\ref{sec:: 3-Modelling}, we analyze the impact of FPVs on MZIs (device level) and array of cascaded MZIs (network level) in SPNNs. Section~\ref{sec:: 4-Optimization} presents our MZI design optimization solution to design devices that are tolerant to different FPVs. Section~\ref{sec:: 5-Results} presents simulation results that highlight how the optimized MZIs can improve the accuracy of the unitary transformation (at the layer level) and the inferencing accuracy (at the system level), in the presence of FPVs. Last, we draw conclusions in Section~\ref{sec:: 6-Conclusion}.
\section{Background and Prior Related Work} 
\label{sec:: 2-Background}
\begin{figure}[t]
\centering
\includegraphics[width=3.5in]{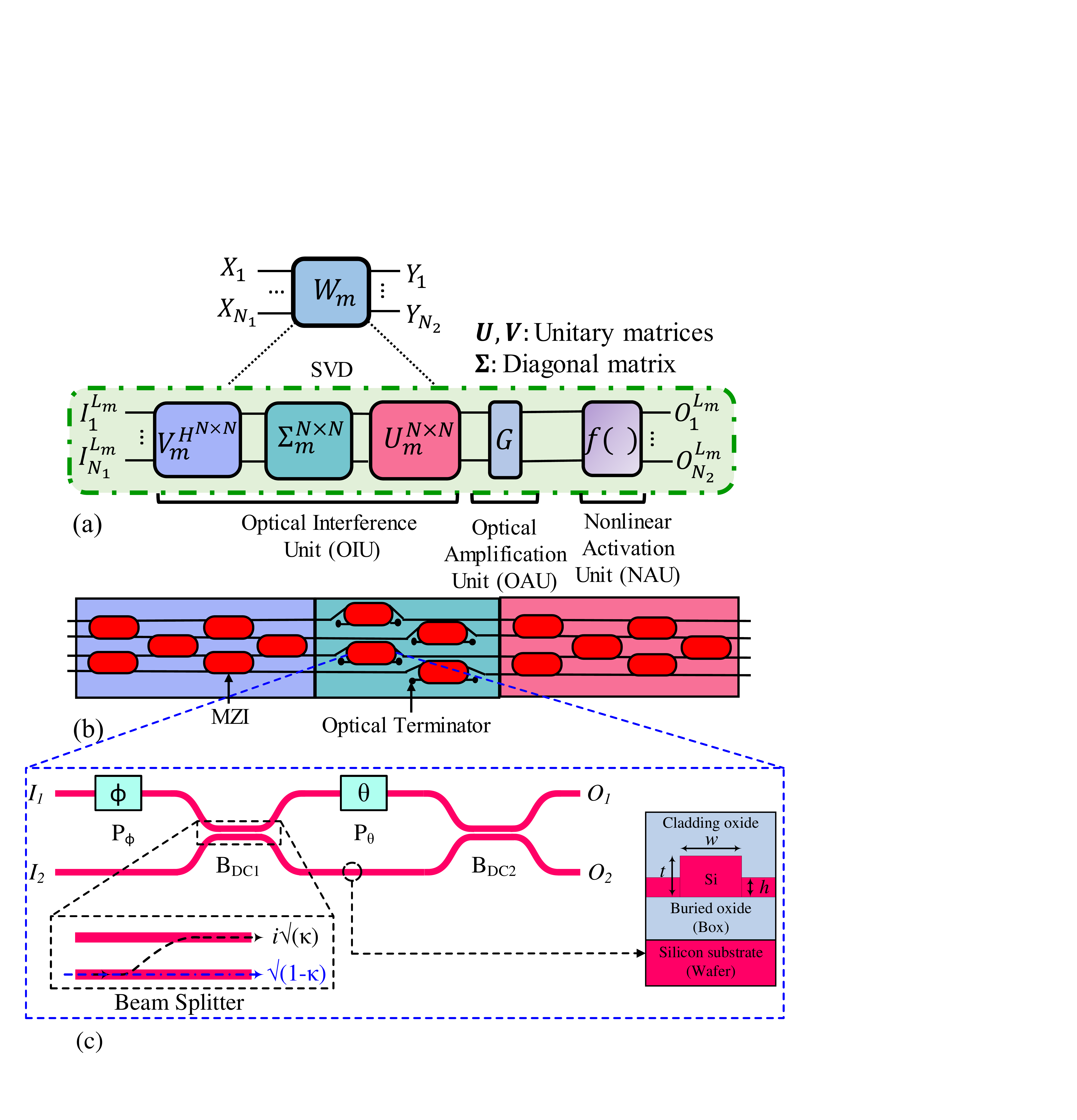}
  \caption{(a) Overview of singular value decomposition (SVD) of a weight matrix related to a fully connected layer ($L_m$) with $N_1$ as the number of input ports and $N_2$ as the number of output ports. (b) An optical-interference unit (OIU). (c) A 2$\times$2 MZI structure with two integrated phase shifters ($\theta$ and $\phi$) and two beam splitters based on directional couplers (DCs).  }\label{Fig:: MZI-Structure}
  \vspace{-1em}
\end{figure}
This section summarizes fundamentals of MZIs and SPNNs designed based on MZIs. Also, it presents some preliminary models to help understand the impact of FPVs on silicon photonic waveguides, and reviews some prior related work.

\subsection{Mach--Zehnder Interferometer (MZI)}
A 2$\times$2 MZI in a SPNN consists of two tunable phase shifters ($\phi$ and $\theta$) on the upper arm and two 50:50 beam splitters as shown in Fig.~\ref{Fig:: MZI-Structure}(c). The phase shifters are used to apply configurable phase shifts and obtain varying degrees of interference between the optical signals traversing the MZI arms. In SPNNs, they are generally implemented using thermal micro-heaters where the effective index of the underlying waveguide changes with temperature (i.e., due to thermo-optic effect of silicon), hence altering the phase of the optical signal. Moreover, 2$\times$2 beam splitters can be designed using directional couplers (DCs) where a fraction of the optical signal (defined as $\kappa$) at an input port is transmitted to an output port, and the remaining signal is coupled to the other output port, as shown in the inset of Fig.~\ref{Fig:: MZI-Structure}(c). The ideal transfer matrix ($T_{MZI}$) for an MZI with two phase shifters ($\theta$ and $\phi$) and two 50:50 beam splitters can be defined as \cite{fang2019design} (Fig. \ref{Fig:: MZI-Structure}(c)):
\begin{align*} 
    T_{MZI}(\theta,\phi) &= B_{DC2} \cdot P_{\theta} \cdot B_{DC1} \cdot P_{\phi} \\
    &= \begin{pmatrix}
      \frac{e^{i\phi}}{2}(e^{i\theta} - 1) & \frac{i}{2}(e^{i\theta} + 1)  \\
      \frac{ie^{i\phi}}{2}(e^{i\theta} + 1) & -\frac{1}{2}(e^{i\theta} - 1)
      \end{pmatrix}, \tag{1}\label{Eq:: Ideal_Transfer_Matrix}
\end{align*}
where $B_{DC1, DC2}$ is the ideal 50:50 beam-splitter transfer matrix and $P_{\phi, \theta}$ is the ideal phase-shifter transfer matrix.

\subsection{Coherent SPNNs based on MZIs}\label{SPNN_back}

Compared to non-coherent SPNNs, coherent SPNNs---considered in this paper---use a single wavelength and MZI devices, in which the adjusted phase shifts in the phase shifters denote the dynamic weight parameters. Fig.~\ref{Fig:: MZI-Structure}(a) shows an example of a coherent SPNN. A fully connected layer in a deep neural network ($L_m$) can be realized with $n_m$ neurons. Each layer performs a linear matrix-vector multiplication and accumulation (MAC) and passes the outputs to the next layer. The output vector of $L_m$ can be mathematically modeled as $O_{m}^{n_m \times 1} = f_m(W_m \times O_{m-1}^{n_{m-1} \times 1})$. Here, $f_m$ is a non-linear activation function (performed by non-linear activation unit (NAU) in Fig.~\ref{Fig:: MZI-Structure}(a)) of $L_m$, and $W_m$ is the corresponding weight matrix of $L_m$. Given a weight matrix $W_m$, which can be obtained by training the network and mapped to MZIs using singular value decomposition (SVD), each weight matrix $W_m$ can be written as $W_m = U_m^{n_m \times n_m}\Sigma_m^{n_m \times n_m}V_m^{H, n_m \times n_m}$. Here, $U_m^{n_m \times n_m}$ and $V_m^{H, n_m \times n_m}$ are the unitary matrices with dimension of $n_m \times n_m$, and $\Sigma_m^{n_m \times n_m}$ is a diagonal matrix with dimension of $n_m \times n_m$. Also,  $V_m^{H, n_m \times n_m}$ is Hermitian-transpose of $V_m^{n_m \times n_m}$. A unitary matrix can be realized by using a cascaded array of 2$\times$2 MZIs. As shown in Figs.~\ref{Fig:: MZI-Structure}(a) and \ref{Fig:: MZI-Structure}(b), this unit is called the optical-interference unit (OIU) and is responsible for performing the MAC operation. 

Several approaches have been proposed to derive the architecture of MZI arrays to perform MAC operations in the optical domain \cite{Reck,Clements:16,Shokraneh:20_Diamond}. Out of these, the Clements design, due its symmetric nature, has the lowest optical loss and footprint. Therefore, we use the Clements design \cite{Clements:16} to transform each unitary matrix ($U_m^{n_m \times n_m}$ and $V_m^{H, n_m \times n_m}$) to a cascaded MZI array with a specific phase setting per MZI in the network. In the Clements method, each unitary matrix will be mapped to a cascaded MZI array with a total number of $\frac{N(N-1)}{2}$ MZIs, where $N$ is the size of the designated unitary matrix. Note that the diagonal matrix ($\Sigma_m^{n_m \times n_m}$) can be realized with an array of MZIs with one input and one output terminated. The optical-amplification unit (OAU), which is required to obtain arbitrary diagonal matrix, in Fig.~\ref{Fig:: MZI-Structure}(a) can be realized realized using semiconductor optical amplifiers (SOAs).
\begin{table}[t]
  \centering
  \caption{Parameters used to generate FPV maps.}
  \begin{tabular}{|c|c|c|}
    \hline
    \rowcolor{Gray}
    \textbf{Design Parameter} & \textbf{Correlation Length ($l$)} & \textbf{Standard Deviation ($\sigma$)}\\
    \hline
    {Waveguide width} & {1~mm and 100~$\mu$m} & $\sigma_w=$~5~nm \\
    \hline
    {SOI thickness} & {1~mm and 100~$\mu$m} & $\sigma_t=$~2~nm \\
    \hline
  \end{tabular}
  \label{tab:: Std_Corr}
  \vspace{-1em}
\end{table}

\subsection{Fabrication-Process Variations (FPVs)}
FPVs in silicon photonics originate in optical-lithography process imperfections, contributing to changes in the waveguide width and SOI thickness. Such changes deviate the effective index ($n_{eff}$), and in turn the propagation constant ($\beta$), in a waveguide, which determines the optical phase of the signal traversing the waveguide. Considering Fig.~\ref{Fig:: MZI-Structure}(c) and as an example, the effective index ($n_{eff}$) in a strip waveguide (with $h=$~0) depends on the optical wavelength and the critical dimensions of the waveguide (i.e., width ($w$) and thickness ($t$) in Fig.~\ref{Fig:: MZI-Structure}(c))~\cite{nikdast2016chip}. The relation can be defined as:
\setcounter{equation}{1}
\begin{equation}
    n_{eff}(\lambda, w, t) = \left(\frac{\lambda}{2\pi}\right)\beta(\lambda, w, t), 
    \label{Eq:: Efective_Index}
\end{equation}
where $\lambda$ is the optical wavelength. Here, $\beta$ can be defined as $\beta=\psi/L$, where $\psi$ is the single pass phase-shift induced in a waveguide of length $L$. Leveraging \eqref{Eq:: Efective_Index}, propagation constant changes ($\Delta\beta$) in a strip waveguide under FPVs is given by: 
\begin{equation}
    \Delta\beta = \frac{2\pi}{\lambda}\left(\frac{\partial n_{eff}}{\partial w}\rho_w + \frac{\partial n_{eff}}{\partial t}\rho_t\right).
    \label{Eq:: Delta_Beta}
\end{equation}
Here, $\rho_w$ and $\rho_t$ are the variations in the waveguide width and SOI thickness, respectively. Note that, in this paper, we do not consider the variations in $h$ (i.e., slab thickness in ridge waveguides) and $L$ (see Section IV for more details). 

\subsection{Related Work on FPV Analysis in SPNNs}
To alleviate the effect of FPVs in MZIs, thermal actuators are usually used to compensate for phase errors, which 
leads to mutual thermal crosstalk among neighboring waveguides \cite{milanizadeh2019canceling}. The work in \cite{banerjee2021modeling} studied the impact of random phase noise due to FPVs and thermal crosstalk in SPNNs by developing a framework that identifies critical components in the network. In \cite{fang2019design}, imprecisions were introduced in SPNNs after software training such that pre-fabrication training tends to be more scalable in terms of network size and volume. This helps designing precise and cost effective MZIs, when compared to re-configurable ones, which can be exploited for AI applications to perform matrix multiplication. A method was presented in \cite{zhu2020countering} to counter the impact of both FPVs and thermal effects using modified cost functions during training with added benefits of post-fabrication hardware calibration. The impact of FPVs can also be reduced by minimizing the tuned phase angles in an SPNN; this can be done by leveraging the non-uniqueness of SVD \cite{banerjee2021optimizing} or by pruning redundant phase angles \cite{banerjee2021pruning,banerjee2021champ}. All these methods focus on mitigating deviations in SPNNs post-fabrication by either using thermal actuators or by post-fabrication training methods to compensate for any additionally introduced phase noise.

In this work, we model the impact of FPVs in the waveguide width and SOI thickness in coherent SPNNs for the first time. We also propose optimizing the design of the underlying MZIs in the network. We show that by exploring and optimizing MZI device physical-level design, we can improve the relative-variation distance ($RVD$) in SPNNs, which quantifies the deviation between the intended unitary matrix and the deviated unitary matrix, to enhance the overall inferencing accuracy. 

\section{Modeling FPVs in Coherent SPNNs} 
\label{sec:: 3-Modelling}
This section presents a detailed bottom-up analysis of the impact of FPVs in the waveguide width and SOI thickness at the device level (i.e., MZI devices) and network level (OIU in Fig.~\ref{Fig:: MZI-Structure}(b)) in coherent SPNNs. We show the impact of FPVs on SPNN inferencing accuracy (system level) in Section V. 
\begin{figure}[t]
\centering
  \includegraphics[width=3.5in]{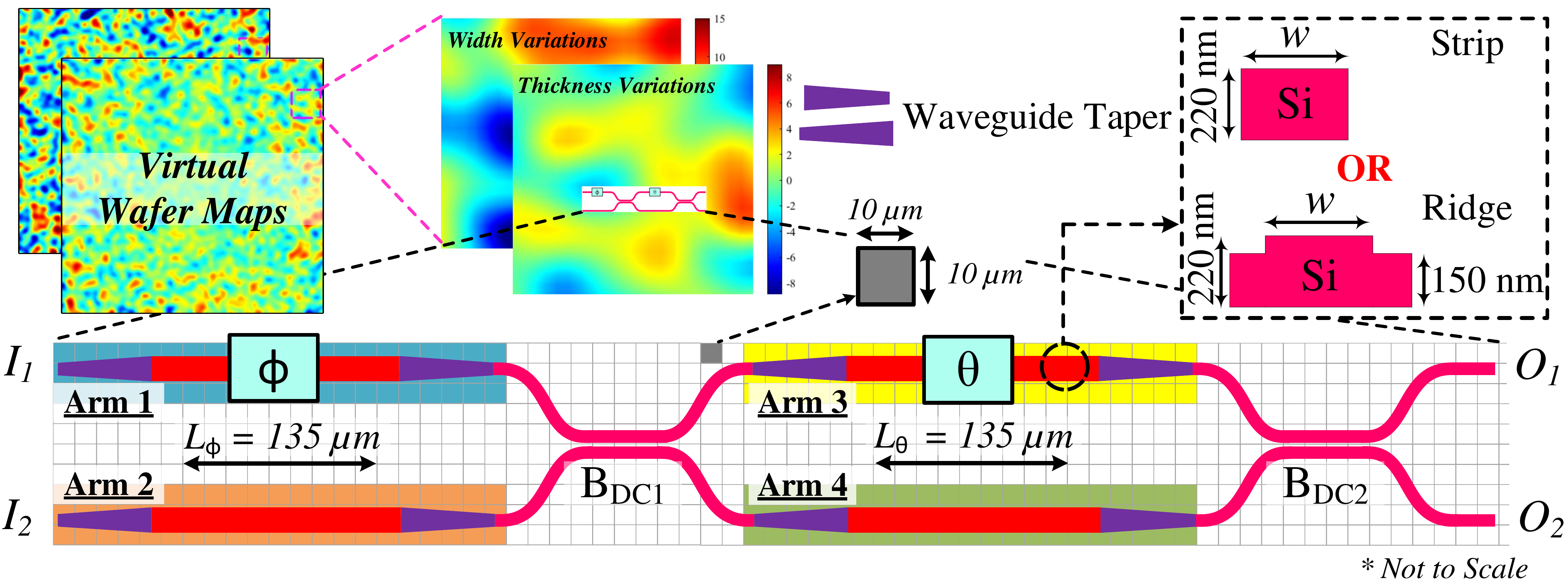}
  \caption{An MZI device structure with waveguide tapers mapped to a FPV map (top), developed based on our work in \cite{mirza2021silicon}, with a mesh size of 10~$\mu$m. The MZI can use strip waveguides or shallow-etched ridge waveguides, both with the SOI thickens of 220~nm and varying waveguide width ($w$) on each arm.}
  \label{Fig:: Variations In MZI}
  \vspace{-1em}
\end{figure}

\subsection{Device Level: MZI Performance under FPVs}
To study the impact of FPVs on MZI devices, we should first model FPVs in silicon photonics and explore how MZI devices experience such FPVs. In our prior work \cite{mirza2021silicon}, we have developed realistic wafer variation maps that replicate radial-variation effects and correlation among different variations---both of which are critical to realistically model FPVs in silicon photonics---in SOI wafers. Such maps were developed based on mean, standard deviation ($\sigma$), and correlation lengths ($l$) experimentally characterized in collaboration with CEA-Leti in \cite{mirza2021silicon}. Table~\ref{tab:: Std_Corr} summarizes different parameters considered to generate FPV maps for our calculations in this work. Fig.~\ref{Fig:: Variations In MZI}-top shows examples of wafer and die maps generated using our in-house FPV simulator with a resolution of 10$\times$10~$\mu$m (i.e., the mesh size in the map---see Fig.~\ref{Fig:: Variations In MZI}).

MZIs are bulky devices (e.g., $\approx$340~$\mu$m in length \cite{shokraneh2020theoretical}), and hence every section of the device will experience slightly different FPVs (see Fig.~\ref{Fig:: Variations In MZI}). Such a difference changes with the correlation length in FPV maps. As a result, it is critical to analyze the impact of the non-uniformity of FPVs in MZI devices. Considering the MZI shown in Fig.~\ref{Fig:: Variations In MZI} with its four arms labeled (Arm1--Arm4), we average the width and SOI thickness variations observed on each MZI arm, separately over the section colored on each arm. In this paper, we consider the state-of-the-art MZI designed in \cite{shokraneh2020theoretical} with the nominal design parameters $t=$~220~nm and $w=$~470~$\mu$m (strip waveguide). The total length of a phase shifter ($L_{\phi}$ and $L_{\theta}$ in Fig.~\ref{Fig:: Variations In MZI}) is $\approx$135$~\mu$m in \cite{shokraneh2020theoretical}. We consider $\approx$10$~\mu$m long DCs with a 200~nm gap to obtain 50:50 splitting ratio in the MZI \cite{nikdast2016chip}. Note that the analyses proposed in this section are independent of the example MZI considered.

Considering the MZI in Fig.~\ref{Fig:: Variations In MZI}, the optical signals traversing the two opposite arms of the device---before the input DC (i.e., Arms 1 and 2) and output DC (i.e., Arms 3 and 4)---should only experience the desired phase change $\phi$ or $\theta$, adjusted after the network training. Note that the phase shifters are integrated on the top of the silicon waveguides. However, due to the impact of non-uniform FPVs on each individual arm, the optical signals experience some undesired phase changes. Assuming that each arm's length ($L_1=L_2=L_\phi$ and $L_3=L_4=L_\theta$) does not undergo any variations, the optical phase difference observed at the input of DC1 ($\Delta\Phi_{DC1}$) and DC2 ($\Delta\Phi_{DC2}$) and between the two optical signals traversing the opposite arms is:
\begin{align*}
    \Delta\Phi_{DC1}=\phi + \left|\Delta\beta_1 L_1 - \Delta\beta_2 L_2\right|,\tag{4a}\label{Eq:: Delta_Phi-a}\\
    \Delta\Phi_{DC2}=\theta + \left|\Delta\beta_3 L_3 - \Delta\beta_4 L_4\right|.\tag{4b}\label{Eq:: Delta_Phi-b}
\end{align*}
Here, $\Delta\beta_1$, $\Delta\beta_2$, $\Delta\beta_3$, and $\Delta\beta_4$ are, respectively, the propagation constant changes on MZI's Arms 1--4, which can be calculated using \eqref{Eq:: Delta_Beta}. 



In addition to signal phase noise, FPVs can impact the splitting ratio, which should ideally be 50:50, in the input and output directional couplers (DC1 and DC2 in Fig.~\ref{Fig:: Variations In MZI}). Our prior work in \cite{banerjee2021modeling} showed that variations in the DC splitting ratio can lower SPNN inferencing accuracy by 30\%. The splitting ratio in a DC is determined by the cross-over coupling coefficient ($\kappa$) \cite{nikdast2016chip}, which itself is a function of the optical wavelength and critical dimensions in the DC. Employing supermode theory \cite{nikdast2016chip}, the impact of FPV-induced changes in the cross-over coupling ($\Delta\kappa$) in a DC can be modeled as:
\begin{equation}
    \Delta\kappa = I \sin^2 (\frac{\pi L_{DC} \Delta n_{DC}}{\lambda_0}),\tag{5}
    \label{Eq:: Cross-Over Coupling Coefficient}
\end{equation}
where $I$ is the input electric field, $L_{DC}$ is the length of DC ($L_{DC}=$~10~$\mu$m in this paper), and $\lambda_0$ is the central wavelength, considered as 1550~nm in this paper. Also, $\Delta n_{DC}$ denotes the changes in the effective index in the DC, which can be calculated based on~\eqref{Eq:: Efective_Index} and supermode analysis in \cite{nikdast2016chip}.

Leveraging \eqref{Eq:: Delta_Beta}, (4), and \eqref{Eq:: Cross-Over Coupling Coefficient}, the MZI transfer matrix in \eqref{Eq:: Ideal_Transfer_Matrix} can be updated to take into consideration the impact of FPVs on MZI arms ($\Delta\beta$) (i.e., undesired signal phase noises) and those on DCs ($\Delta\kappa$) (i.e., splitting ratio deviations):

\begin{align*} 
    &T'_{MZI}(\theta,\phi) = B'_{DC2} \cdot P'_{\theta} \cdot B'_{DC1} \cdot P'_{\phi} \\
    &= \left(\begin{smallmatrix}
      \sqrt{1-(\kappa_2 + \Delta\kappa_2)} & i\sqrt{\kappa_2 + \Delta\kappa_2}  \\
      i\sqrt{\kappa_2 +\Delta\kappa_2} & \sqrt{1-(\kappa_2 + \Delta\kappa_2)}
      \end{smallmatrix}\right)
    \cdot 
    \left(\begin{smallmatrix}
      e^{i(\theta+\Delta\beta_3 L_3)} & 0\\
      0 & e^{i\Delta\beta_4 L_4}
      \end{smallmatrix}\right)
    \cdot \\
     &\left(\begin{smallmatrix}
      \sqrt{1-(\kappa_1 + \Delta\kappa_1)} & i\sqrt{\kappa_1 +\Delta\kappa_1}  \\
      i\sqrt{\kappa_1+ \Delta\kappa_1} & \sqrt{1-(\kappa_1 + \Delta\kappa_1)}
      \end{smallmatrix}\right)
    \cdot 
     \left(\begin{smallmatrix}
      e^{i(\phi+\Delta\beta_1 L_1)} & 0\\
      0 & e^{i\Delta\beta_2 L_2}
      \end{smallmatrix}\right), \tag{6}\label{Eq:: Transfer_Matrix}
\end{align*}
where, as shown in Fig.~\ref{Fig:: Variations In MZI}, $L_1=L_2=L_\phi$ and $L_3=L_4=L_\theta$, and $L_\phi=L_\theta=$~135~$\mu$m considered as an example from \cite{shokraneh2020theoretical} in this paper. Leveraging (\ref{Eq:: Transfer_Matrix}), we can capture the impact of non-uniform variations in the waveguide width and SOI thickness in any MZI design in coherent SPNNs.

\subsection{Network Level: OIU Performance under FPVs}

Here, we model the impact of FPVs on the performance of an OIU, shown in Fig.~\ref{Fig:: MZI-Structure}(b). Note that in this paper we do not consider the impact of FPVs in the OAUs and NAUs (see Fig.~\ref{Fig:: MZI-Structure}(a)) as they are often implemented either electronically or opto-electronically \cite{PourFard:20,cheng2020silicon}. Recall from Section \ref{SPNN_back} that, given a weight matrix $W = U \Sigma V^{H}$, the matrices $U$, $\Sigma$, and $V$ can be decomposed to $\theta$ and $\phi$ phase values on each MZI in an OIU using Clements decomposition \cite{Clements:16}. Under FPVs, the transfer matrix of each MZI in the OIU will deviate in a manner that can be calculated using \eqref{Eq:: Transfer_Matrix}. To analyze the impact of such variations at the network level, we use relative-variation distance ($RVD$). $RVD$ determines the deviation between an intended  matrix and a deviated  matrix \cite{banerjee2021modeling}. In our prior work \cite{banerjee2021modeling}, we observed a strong correlation between network-level $RVD$ and system-level inferencing accuracy in SPNNs (we will discuss this in Section V). $RVD$ can be defined as:
\begin{equation}
    RVD(W,\overline{W}) = \frac{\Sigma_m\Sigma_n |W^{m,n}-\overline{W}^{m,n}|}{\Sigma_m\Sigma_n |W^{m,n}|},
    \tag{7}\label{Eq:: RVD}
\end{equation}
where $|.|$ denotes the absolute value of a complex number. $W$ is the nominal weight matrix and $\overline{W}$ is the deviated weight matrix under FPVs related to a fully connected layer. $W^{m,n}$ denotes the element at the $m$\textsuperscript{th} row and $n$\textsuperscript{th} column of $W$. Each MZI in the OIU has a unique impact on each element of the weight matrix. Accordingly, variations related to each MZI in the network have a unique effect on the overall $RVD$. Higher $RVD$ means the actual weight matrix is more deviated from the intended one, which can be interpreted as observing a lower inferencing accuracy at the system level. 

To compute $\overline{W}$ in \eqref{Eq:: RVD}, we first need to analyze the impact of non-uniform FPVs in the waveguide width and SOI thickness on each individual MZI in an OIU. As the first step, we calculate the total dimension of an OIU based on the length of an individual MZI ($l_{MZI}$) and the gap between its input and output ports ($g_{MZI}$). In this work and as an example, we consider the state-of-the-art MZI design from \cite{shokraneh2020theoretical} with $l_{MZI}=$~340~$\mu$m and $g_{MZI}=$~30~$\mu$m. Accordingly and using the Clements design for the OIU \cite{Clements:16}, the total dimension of the OIU can be calculated. As the second step, we use our in-house FPV map simulator (see Section III-A) to generate a die map that matches the OIU. We then place the OIU on the die map and extract FPV information for each individual MZI in the OIU. Note that each MZI itself experiences different variations (non-uniform variations across a single device) and the FPVs between two different MZIs are also different. All such non-uniformities are considered in our device-level (see Section III-A) and network-level analyses. By capturing FPVs for each individual MZI in an OIU, we can calculate the deviated $\overline{W}$ in \eqref{Eq:: RVD} based on:
\begin{equation}
    \overline{W}_m = \overline{U}_m \times \overline{\Sigma}_m \times \overline{V_{m}^{H}}.\tag{8} \label{SVD_weights}
\end{equation}
Here, $\overline{U}_m$ and $\overline{V_{m}^{H}}$ are the deviated unitary transfer matrices and $\overline{\Sigma}_m$ is the deviated diagonal matrix under FPVs. They can be calculated based on multiplying MZI transfer matrices under FPVs (the model in \eqref{Eq:: Transfer_Matrix}) and in a specific order determined by the Clements design \cite{Clements:16}. For example, for $\overline{U}_m$ we have:
\begin{equation}
\overline{U}_m=D\left(\prod_{(m, n) \in S} T_{MZI,m, n}^{'}\right),
\tag{9}\label{connection}
\end{equation}
where $m$ and $n$ should be calculated based on the mapping method (e.g., Clements \cite{Clements:16}) used to map the weight matrices to cascaded MZI arrays. Also, $S$ is the order of multiplication which again should be determined by the mapping method. Moreover, $D$ is a diagonal matrix with unity magnitude and is not related to the physical placement of MZIs. Similarly, we can calculate $\overline{V_{m}^{H}}$. Although we considered the Clements method for mapping the weights to phase settings of a cascaded MZI array in OIUs, our network-level models in this section can work with any other mapping method.

\begin{figure}[t]
\centering
  \captionsetup[subfigure]{oneside,margin={0cm,0cm}}
    \subfigure[Strip waveguide]{\hspace{-0.1cm}\includegraphics[width=1.74in]{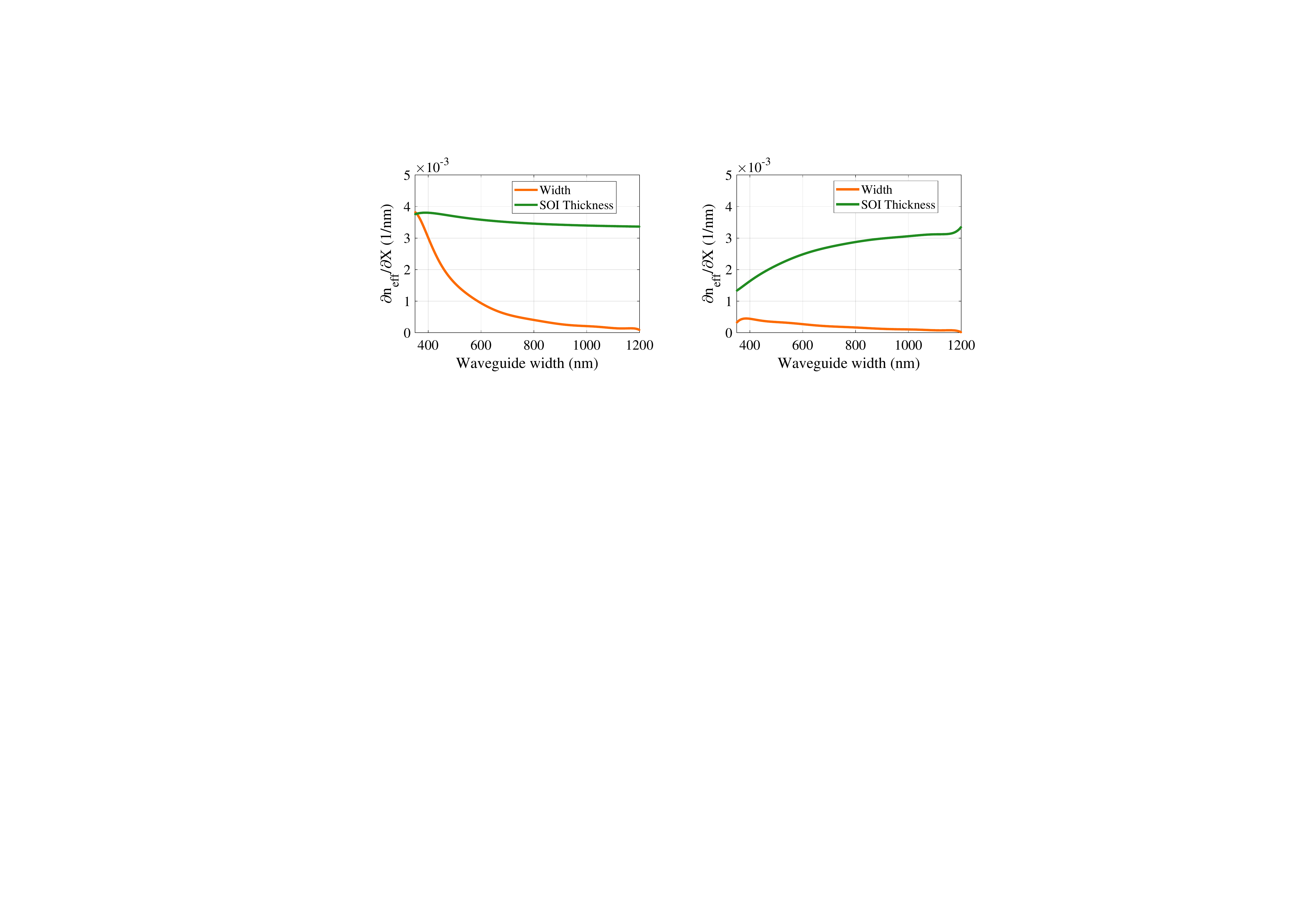}\hspace{0cm}} 
    \captionsetup[subfigure]{oneside,margin={0cm,0cm}}
    \subfigure[Ridge waveguide]{\hspace{0cm}\includegraphics[width=1.74in]{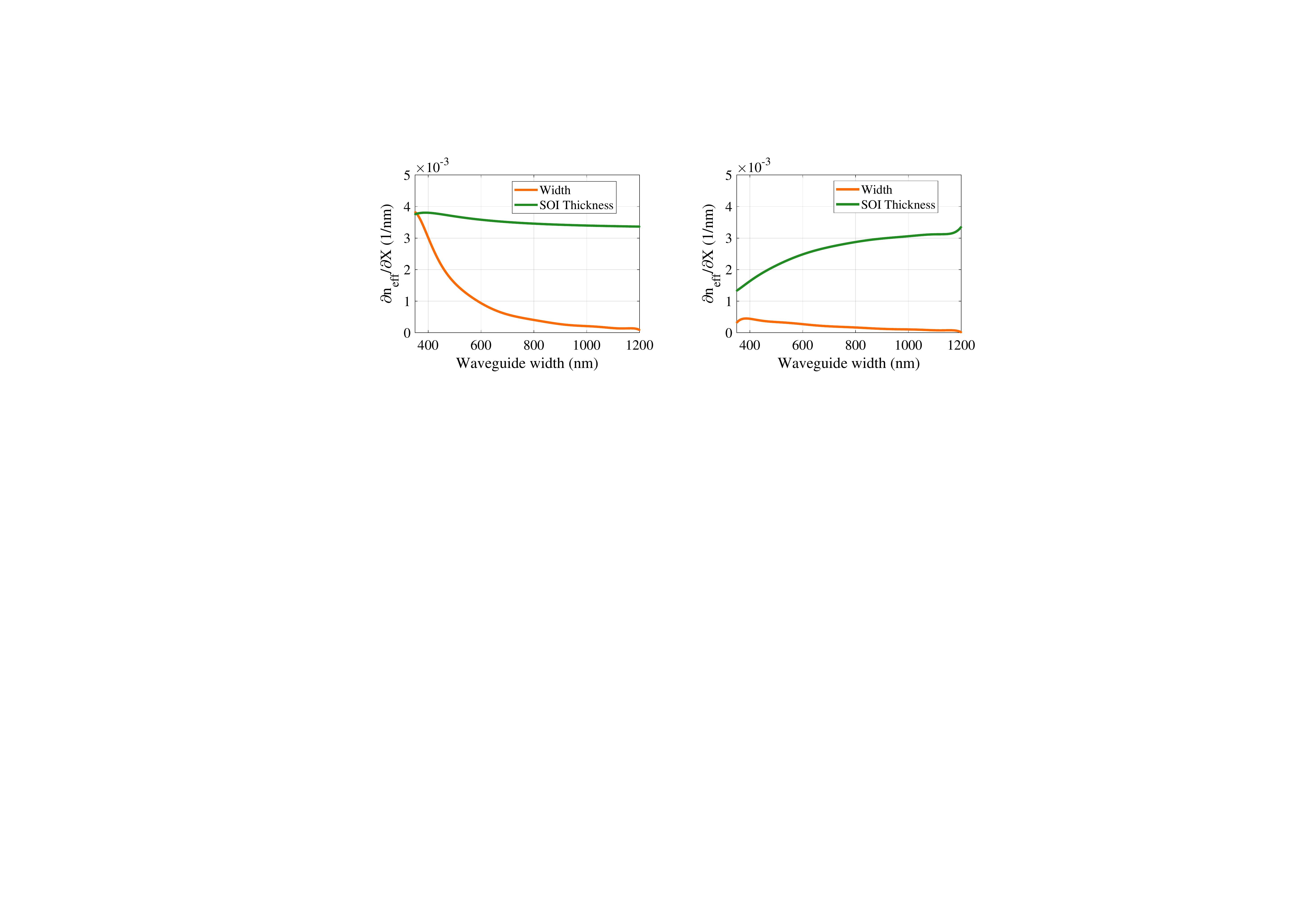}\hspace{0cm}}
  \caption{Rate of changes in waveguide effective index (see the strip and ridge waveguides in Fig.~\ref{Fig:: Variations In MZI}) under FPVs ($\frac{\partial n_{eff}}{\partial X}$), where $X$ shows the design parameter under FPVs, in (a) a strip and (b) a shallow-etched ridge waveguide, when the waveguide width ($w$) increases from 350 to 1200~nm.}\label{Fig:: dneff/dX}
  \vspace{-0.2in}
\end{figure}
\section{SPNN Design Optimization Under FPVs} 
\label{sec:: 4-Optimization}
In this section, we explore the design space of MZI devices under different FPVs to optimize their performance in SPNNs. In particular, we focus on minimizing the impact of FPVs on MZI arms that imposes undesired optical phase noises in the device, leading to faulty matrix-vector multiplication. As discussed in Section~\ref{sec:: 3-Modelling}, FPVs also deviate the splitting ratio in DCs in an MZI. Nevertheless, the design optimization solution in this section assumes ideal DCs. Note that FPV-tolerant DCs can be designed based on the method proposed in \cite{lu2015broadband}. 

Considering (4), one can alleviate the impact of FPV-induced phase noise in an MZI by implying $\left|\Delta\beta_1 L_1 - \Delta\beta_2 L_2\right|\rightarrow 0$ and $\left|\Delta\beta_3 L_3 - \Delta\beta_4 L_4\right|\rightarrow 0$. Accordingly, to obtain a phase-noise-free MZI design, we should have:
\begin{equation*}
    \frac{L_1}{L_2} = \frac{\Delta{\beta2}}{\Delta{\beta1}}~~\textrm{and}~~ \frac{L_3}{L_4} = \frac{\Delta{\beta4}}{\Delta{\beta3}}.\tag{10}
    \label{Eq:: L1/L2} 
\end{equation*}
This indicates that the length ratio between any two opposite arms in the MZI should be inversely proportional to the ratio of the changes in their waveguide propagation constants ($\Delta\beta$), under non-uniform FPVs. In this section and for brevity, we assume $L_1=L_2=L_3=L_4$ and without variations. As a result and based on \eqref{Eq:: L1/L2}, we should minimize $|\Delta\beta_1-\Delta\beta_2|$ and $|\Delta\beta_3-\Delta\beta_4|$ under different FPVs. In other words, we should make sure that the propagation constant changes on the two opposite arms in an MZI are as small as possible (i.e., $\Delta\beta_1\rightarrow0$, $\Delta\beta_2\rightarrow0$, $\Delta\beta_3\rightarrow0$, and $\Delta\beta_4\rightarrow0$), or the propagation constant changes on the two opposite arms are as close as possible (i.e., $\Delta\beta_1\rightarrow\Delta\beta_2$ and $\Delta\beta_3\rightarrow\Delta\beta_4$).

Considering \eqref{Eq:: Delta_Beta}, $\Delta\beta$ is proportional to the rate of changes in the waveguide's effective index under FPVs (i.e., $\frac{\partial n_{eff}}{\partial X}$, where $X$ denotes the design parameter under FPVs: $X=w$ for width and $X=t$ for SOI thickness variations). In our prior work \cite{mirza2021silicon}, we found that as the waveguide width increases, $\frac{\partial n_{eff}}{\partial X}$ decreases, especially under waveguide width variations (i.e., when $X=w$). This is because as the waveguide width increases, a bigger portion of the optical mode is confined in the waveguide core (and the confinement is also stronger), and hence the variations in the waveguide width will create less distortion in the optical mode in the waveguide. Fig.~\ref{Fig:: dneff/dX}(a) shows the rate of changes in the effective index of a strip waveguide with $t=$~220~nm (see Fig.~\ref{Fig:: Variations In MZI}) and when the waveguide width ($w$) changes from 350~nm to 1200~nm, both considered as an example. As can be seen, as the waveguide width increases, $\frac{\partial n_{eff}}{\partial w}$ decreases sharply but $\frac{\partial n_{eff}}{\partial t}$ decreases sightly and stays higher than $\frac{\partial n_{eff}}{\partial w}$ under different waveguide widths. While waveguide width can be changed during the design time, the SOI thickness cannot be changed; this parameter is determined by the host SOI wafer. 
\begin{figure}[t]
\centering
  \includegraphics[width=3.4in]{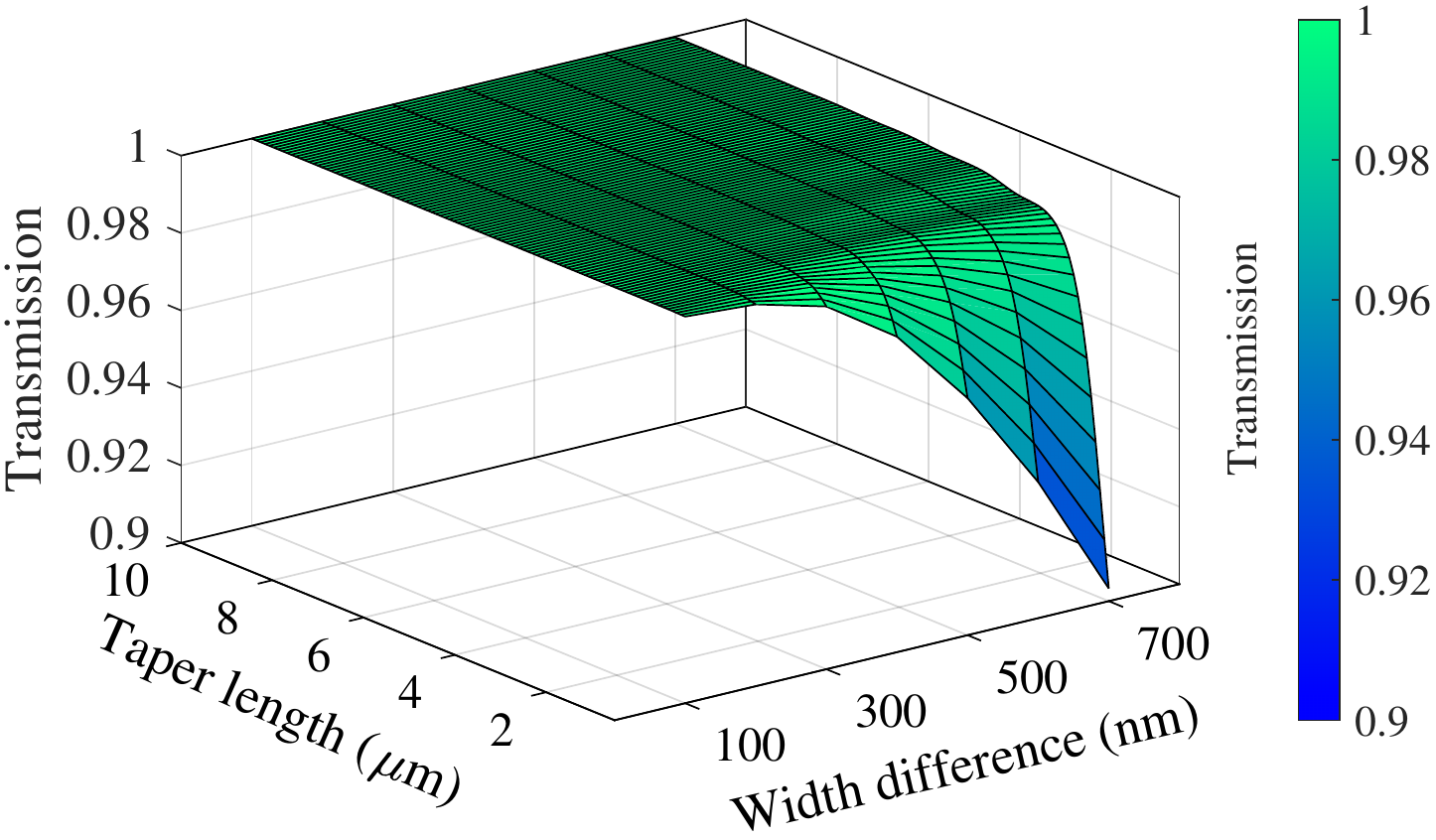}
  \caption{Minimum taper length required to keep the optical transmission between two waveguides of different widths consistent (i.e., at 1 in the figure) and to avoid mode distortion. }
  \label{Fig:: Taper Length}
  \vspace{-0.2in}
\end{figure}

As can be seen from Fig.~\ref{Fig:: dneff/dX}(a), $\frac{\partial n_{eff}}{\partial t}$ is high in strip waveguides and cannot be compensated for by increasing the waveguide width. To address this, we also explore a shallow-etched ridge waveguide with a slab thickness ($h$) of 150~nm, as shown in Fig.~\ref{Fig:: Variations In MZI}. Such a ridge waveguide is common in the design of grating structures \cite{jiang2010slab}. By adding the slab to a strip waveguide (i.e., making it a ridge waveguide), the optical mode is pulled mostly towards the slab region, hence SOI thickness variations should have less impact on the optical mode. Similar to Fig.~\ref{Fig:: dneff/dX}(a), Fig.~\ref{Fig:: dneff/dX}(b) shows the rate of changes in the effective index of the shallow-etched ridge waveguide with $t=$~220~nm and when the waveguide width ($w$) changes from 350~nm to 1200~nm. As can be seen, compared to the strip waveguide, both $\frac{\partial n_{eff}}{\partial w}$ and $\frac{\partial n_{eff}}{\partial t}$ are smaller in the ridge waveguide, as an effect of adding the slab region. Note that we explored different slab thicknesses from 60~nm to 180~nm (results are not shown in the paper), and $h=$~150~nm returned the best results. Furthermore, it is important to note that $\frac{\partial n_{eff}}{\partial h}>0$ in shallow-etched ridge waveguides, but it is much smaller than $\frac{\partial n_{eff}}{\partial t}$. In this paper, we only focus on the critical variations in the waveguide width and SOI thickness.

Considering the results in Fig.~\ref{Fig:: dneff/dX}, designing MZIs with wider strip and shallow-etched ridge waveguides should help minimize the changes in the propagating constant on each MZI arm (see \eqref{Eq:: Delta_Beta} and \eqref{Eq:: L1/L2}). Moreover, to increase the waveguide width, an important design consideration is to include waveguide tapers on the MZI arms as shown in Fig.~\ref{Fig:: Variations In MZI}. Waveguide tapers are essential to avoid optical mode distortion and higher order mode excitation when moving from the nominal waveguide width (i.e., 470~nm in this paper) to a wider waveguide and vice versa \cite{fu2014efficient}. In particular, the waveguide taper length should be long enough to avoid any optical transmission and mode distortion. Using Lumerical MODE \cite{MODE}, we simulated the fundamental mode transmission between two waveguides of different widths in Fig.~\ref{Fig:: Taper Length}. As can be seen, a waveguide taper length of $\approx$1~$\mu$m will be sufficient for every 100~nm width difference between two waveguides of different widths (see Fig.~\ref{Fig:: Variations In MZI}). This helps us calculate the area overhead when we optimize MZIs with wider waveguide widths. 
\begin{figure}[t]
\centering
  \includegraphics[width=0.45\textwidth]{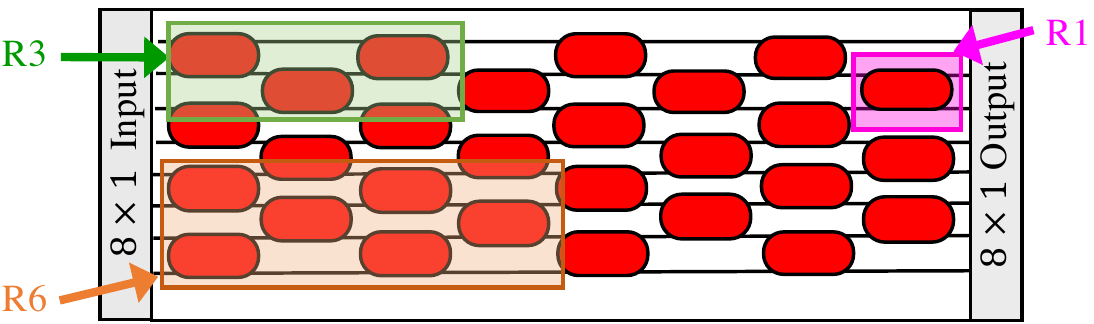}
  \caption{Different region sizes (R1, R3, and R6) and related MZIs in a single 8$\times$8 OIU unit. R12 (not shown in the figure) can be obtained in a similar fashion.}\label{regions}
 \vspace{-0.2in}
\end{figure}

Considering different FPVs in the waveguide width and SOI thickness, which we model based on the wafer map models in \cite{mirza2021silicon}, we consider two scenarios based on which the design of MZIs in an SPNN can be optimized. First, in the \textit{region-based-tolerant MZI design}, we assume a designer may have some \textit{a priori} knowledge of the FPVs. This is a valid assumption as silicon photonics foundries can provide some FPV maps, with different variation data resolutions, to the designers using their fabrication processes. Second, we assume a \textit{worst-case-tolerant MZI design} scenario, where a designer may have very little to no \textit{a priori} knowledge of the FPVs, and hence the MZIs in the SPNN are designed considering the worst-case FPV scenarios (e.g., corner analysis). In such a scenario, we design the worst-case-tolerant MZIs with the largest possible waveguide widths, and equal widths on all the arms, while considering the area overhead in the MZIs. We will further discuss the worst-case-tolerant MZI design and its results in Section V.

For the region-based-tolerant MZI design, we can define different regions of different sizes (i.e., number of MZIs) in the SPNN, an example of which is shown in Fig.~\ref{regions}. Such regions group the MZIs that are spatially close on the die with one another. We assume that the designer has some \textit{a priori} knowledge of the FPVs for all (and not the individual) MZIs grouped in the same region. As a result, the smaller the region size (e.g., R1 with a single MZI in Fig.~\ref{regions}), the more detailed FPV information is available to the designer and vice versa (e.g., R6 with six MZIs in Fig.~\ref{regions}). Accordingly, we consider the average observed variations in a region to design region-based-tolerant MZIs by performing an exhaustive search for the MZI waveguide widths (leveraging findings in Fig.~\ref{Fig:: dneff/dX}) while considering the area overhead in MZIs for the tapers. Here, we might have different waeguide widths (between the search range of 350~nm to 1200~nm) on each MZI arm . 

As we will show in Section V, our region-based-tolerant and worst-case-tolerant MZI designs minimize optical phase noises in MZIs under different FPVs, hence they improve network accuracy in SPNNs. Nevertheless, it is important to note that our device-level design optimization solutions proposed in this section do not aim at completely eliminating, if at all feasible, the impact of FPVs in SPNNs. That being said, our optimization will reduce the impact of FPVs in SPNNs sufficiently so that the overhead and complexity of dynamic calibration techniques to eliminate the impact of such variations in SPNNs will be significantly reduced.

\begin{figure}[t]
 \centering
   \subfigure[$RVD$ for 100 random unitary matrices]{\includegraphics[width=0.42\textwidth]{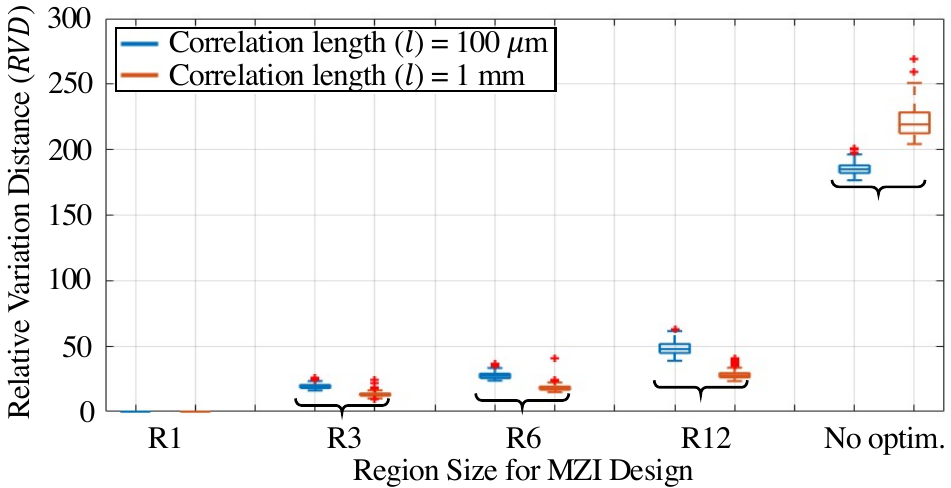}\label{baseline}}\vspace{-0.05in}
  \subfigure[Accuracy and $RVD$ under FPVs (different $\sigma_w$'s and $\sigma_t$'s)]{\includegraphics[width=0.42\textwidth]{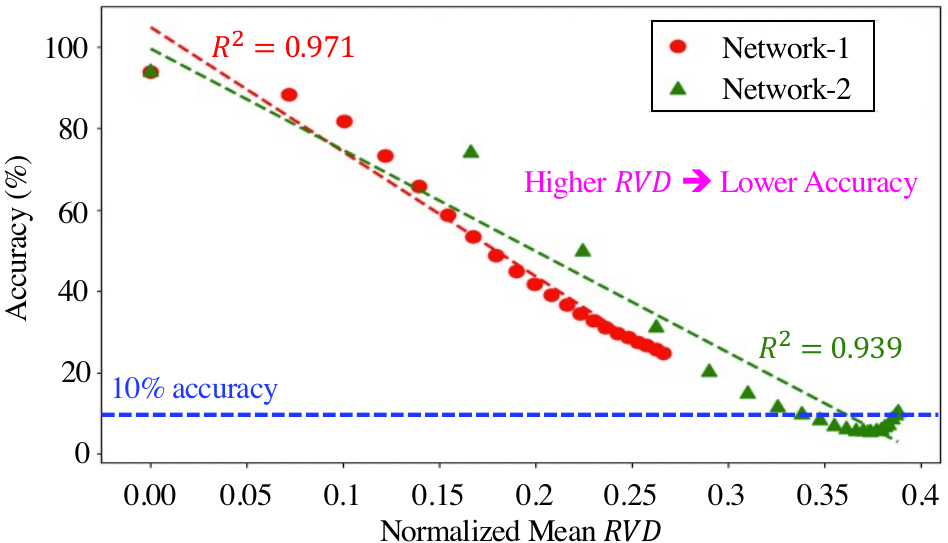}\label{baseline}}\vspace{-0.1in}
  \caption{(a) Statistical analysis of $RVD$ for different region sizes and under FPVs with different correlation lengths. (b) High $R^2$ values denote strong linear correlation between $RVD$ and accuracy. FPVs are based on the parameters in Table~\ref{tab:: Std_Corr}.}
  \label{RVD_regions}
 \vspace{-0.2in}
 \end{figure}
\section{Simulation Results and Discussions} 
\label{sec:: 5-Results}
FPVs lead to undesired optical phase noises, which, in turn, lead to faulty matrix-vector multiplication in the fully connected layers of an SPNN. The sensitivity of a standalone MZI to FPVs depends on its nominal width and thickness (see Fig. \ref{Fig:: dneff/dX}), where only the MZI waveguide width can be altered during design time. Leveraging realistic FPV maps developed based on the method and experimental data in \cite{mirza2021silicon} (see Fig.~\ref{Fig:: Variations In MZI}-top) and the MZI design optimization in Section~\ref{sec:: 4-Optimization}, we explore and optimize the nominal waveguide widths in the MZIs in SPNN case studies considered in this section, to improve their tolerance under FPVs. As a result, unitary matrices realized by connecting such optimized MZIs in the SPNN case studies are also resilient to FPVs. 

Prior to evaluating the impact of the proposed MZI optimization on SPNN accuracy under FPVs, let us explore whether such an optimization is independent of the model and the dataset. To examine this, Fig. \ref{RVD_regions}(a) considers 100 randomly generated 16$\times$16 unitary matrices---each of which belongs to a different weight matrix---and presents a box plot of the distribution of $RVD$'s between the nominal and the deviated unitary matrices under FPVs, and for different region sizes. Recall from Section IV that to design the region-based-tolerant MZIs, we take the mean variations affecting a region on the FPV map with 1, 3, 6, or 12 MZIs into consideration (see Fig. \ref{regions}). All the MZIs in a particular region are replaced with the optimized region-based-tolerant MZI, designed using shallow-etched ridge waveguides (see Figs.~\ref{Fig:: Variations In MZI} and \ref{Fig:: dneff/dX}(b)). Observe that in all cases, the mean $RVD$ is significantly reduced when optimized MZIs are used. In particular, the interquartile ranges (IQRs) in the box plots are consistent among all the unitary matrices in a region: this shows that the proposed optimization is effective independent of the considered unitary matrix. Also, the mean $RVD$ decreases with decreasing region size (i.e., when a designer has access to more detailed FPV data from the foundry), and it is the highest when MZIs are not optimized. 



To explore the impact of our proposed MZI design optimization in Section IV on SPNN accuracy, we consider a case study of two fully connected SPNNs with different footprint (see Table \ref{tab:spnn_arch}) trained on the MNIST dataset. To compress the 28$\times$28~$=$~784 dimensional feature vector in the MNIST dataset, we take the shifted fast Fourier transform of each image. The compressed 16-dimensional feature vector for Network 1 is then obtained by considering the values within the 4$\times$4 region at the center of the frequency spectrum. Similarly, for the larger Network 2, we use a 64-dimensional feature vector by considering the 8$\times$8 region at the center of the frequency spectrum. \textcolor{black}{In Fig. \ref{RVD_regions}(b), we show the linear correlation between the accuracy and the mean $RVD$, averaged over the 6 unitary matrices and normalized over the number of phase shifters in each network. Given such a strong linear correlation, it is expected that the proposed method should improve the accuracy of all SPNNs under FPVs, irrespective of the nominal phase angles.}

\begin{figure}[t]
 \centering
 \captionsetup[subfigure]{aboveskip=10pt} 
   \subfigure[Network-1, correlation length = 100 $\mu$m]{\includegraphics[width=0.49\textwidth]{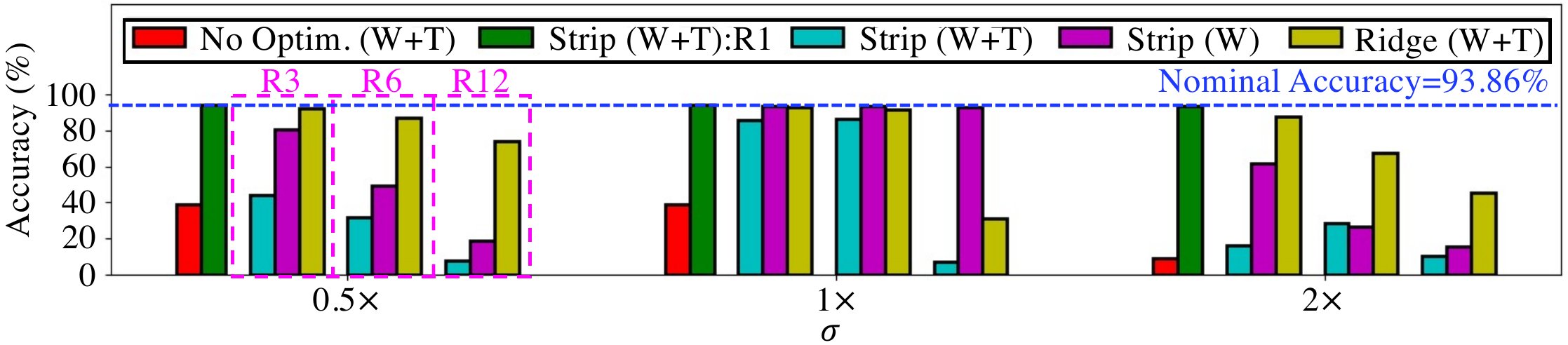}\label{baseline}}\vspace{-0.05in}
  \subfigure[Network-1, correlation length = 1 mm]{\includegraphics[width=0.49\textwidth]{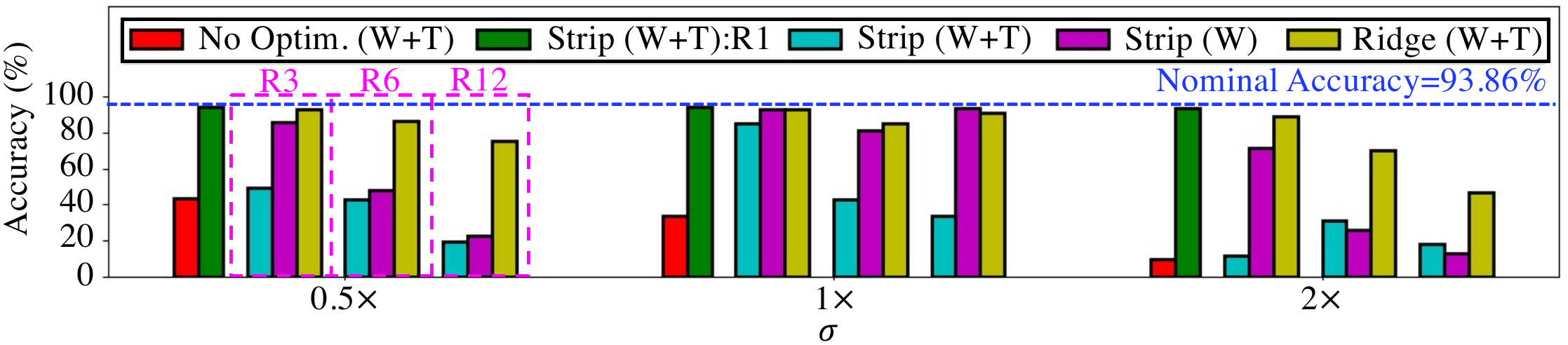} \label{ltpplots}}\vspace{-0.05in}
  \subfigure[Network-2, correlation length = 100 $\mu$m]{\includegraphics[width=0.49\textwidth]{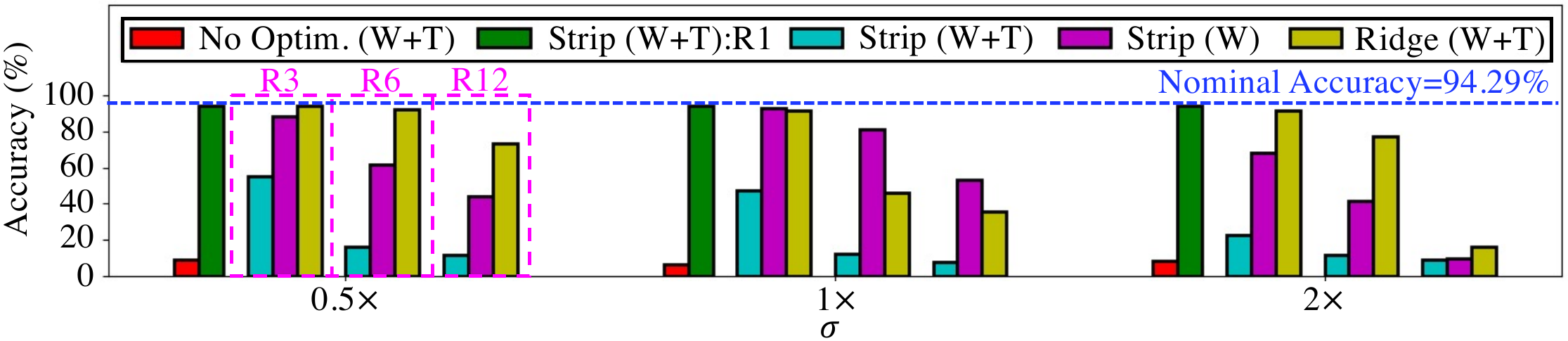}\label{baseline}}\vspace{-0.05in}
  \subfigure[Network-2, correlation length = 1 mm]{\includegraphics[width=0.49\textwidth]{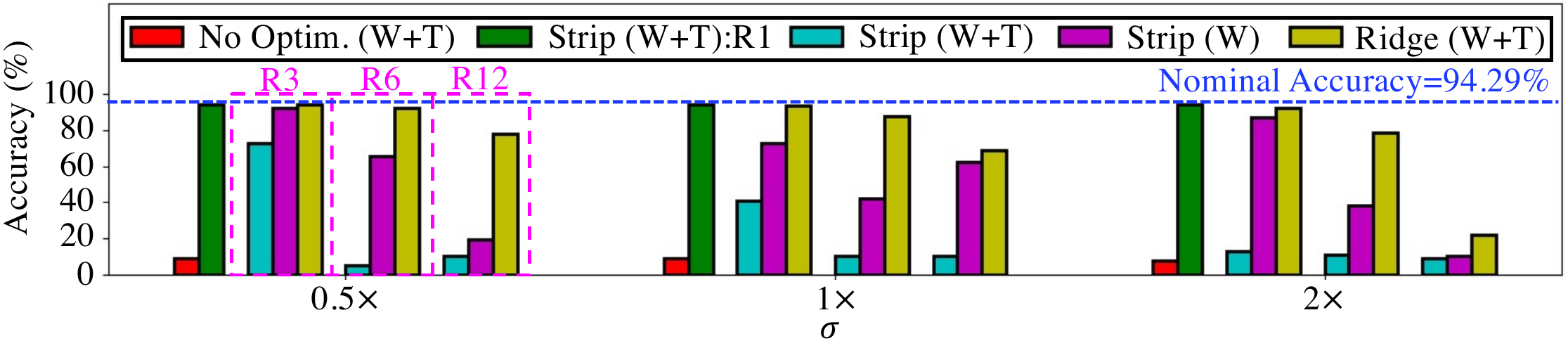}\label{baseline}}\vspace{-0.1in}
  \caption{Accuracy of two SPNNs under FPVs in width and SOI thickness before and after optimization using the proposed method. W: width variation and T: SOI thickness variation.}
  \label{accuracysims}
 \vspace{-0.2in}
 \end{figure}

\begin{table}[t]
    \centering
    \caption{Architectures of the SPNNs considered. FC(x,y): Fully connected layer with x inputs and y outputs, SP: Softplus activation, LSM: LogSoftMax activation, PhS: phase shifters.}
    \begin{tabular}{|c|c|c|c|}
    \hline
    \rowcolor{Gray}
         Model & Architecture & \# PhS  \\
         \hline
         \textbf{Network-1} & FC(16,16)-SP-FC(16,16)-SP-FC(16,10)-LSM & 1380 \\
         \textbf{Network-2} & FC(64,64)-SP-FC(64,64)-SP-FC(64,10)-LSM & 20,580 \\
         \hline
    \end{tabular}
    
    \label{tab:spnn_arch}
    \vspace{-1em}
\end{table}

By applying realistic FPV maps---generated using our in-house simulation \cite{mirza2021silicon} and considering parameters in Table~\ref{tab:: Std_Corr}---to Network-1 and Network-2, Fig. \ref{accuracysims} shows the inferencing accuracy in each network with conventional MZIs (No Optim.) and optimized region-based-tolerant MZIs, which can have different waveguide widths on each arm. In each plot, we consider three different standard deviations for the waveguide width and SOI thickness variations that are 0.5$\times$, 1$\times$, and 2$\times$ the expected standard deviations ($\sigma_w$, $\sigma_t$) in Table~\ref{tab:: Std_Corr}. Moreover, the variation maps are generated for two correlation lengths of 100~$\mu$m and 1~mm. As can be seen from Figs.~\ref{accuracysims}(a)--(d), with no MZI optimization (red bar in the figures), the SPNN accuracy is always the least. Considering optimized region-based-tolerant MZI design with strip waveguides and  R1 (i.e., when regions include a single MZI---see Fig.~\ref{regions}), the network accuracy in Figs.~\ref{accuracysims}(a)--(d) is almost the same as the nominal accuracy. This is because each individual MZI has been optimized considering its exact FPV profile. This is for an ideal case when a designer have full access to variation data affecting each MZI in the network, so it may be impractical in most cases (results for R1 are shown here to indicate the efficiency of our optimization in such rare cases). 

Considering more realistic and practical region sizes (R3, R6, and R12) and a region-based-tolerant MZI design using strip waveguides, Figs.~\ref{accuracysims}(a)--(d) show the SPNN accuracy when considering (i) both waveguide width and SOI thickness variations (W+T; blue bar), and (ii) when only width variations exist (W; magenta bar). As can be observed, with (i), the optimized MZI can help retrieve some accuracy, which also decreases as the region size increases. But overall, the gain in accuracy in this case is small. This is due to the fact that optimized MZIs designed using strip waveguides are not sufficiently tolerant to thickness variations (see Fig.~\ref{Fig:: dneff/dX}(a)). Considering (ii), which is for cases when SOI thickness variations are negligible (e.g., through SOI thickness uniformity improvement \cite{6053719}), the MZI designed using strip waveguides achieves acceptable accuracy improvements (and higher than (i)) in both the networks. 

Finally, Figs.~\ref{accuracysims}(a)--(d) show the network accuracies when optimized region-based-tolerant MZIs designed using shallow-etched ridge waveguides are considered, under the presence of both waveguide width and SOI thickness variations. As can be observed, such MZIs perform better compared to the strip-waveguide-based MZIs discussed above. In fact, at 0.5$\times\sigma$ with R3 and R6, 1$\times\sigma$ with R3, and 2$\times\sigma$ with R3, ridge MZIs almost entirely recover the nominal accuracy under width and SOI thickness variations. This shows that ridge MZIs optimized using our method can effectively make SPNNs tolerant to expected levels of FPVs. Another important observation is that how the network accuracy results in Figs.~\ref{accuracysims}(a)--(d) seem to change across different region sizes and FPV correlation lengths. In fact, when the correlation length in variations is shorter, variations tend to "change more" within a region with a given size, and as the region size increases, they change even more across the region. This imposes a higher error for the region-based-tolerant MZIs designed per region. As a result, the accuracy results are a bit lower in Figs.~\ref{accuracysims}(a) and \ref{accuracysims}(c) compared to those in Figs.~\ref{accuracysims}(b) and \ref{accuracysims}(d).

To assess SPNN accuracy using the optimized worst-case-tolerant MZI design (see Section IV), we consider, as an example, Network-2 with FPVs of different correlation lengths and standard deviations in Table~\ref{tab:: Std_Corr}. In this experiment, we assume no \textit{a priori} FPV knowledge and using strip waveguides in the design of the optimized worst-case-tolerant MZIs while considering waveguide width variations only. The worst-case-tolerant MZI is optimized by widening all the MZI arms together---i.e., MZI arms all have the same width after optimization---while considering the resulting area overhead due to the required waveguide tapers (see Figs. \ref{Fig:: Variations In MZI} and \ref{Fig:: dneff/dX}(a)). Results for this experiment (before and after the optimization) are shown in Table~\ref{Tab:: Phase Noise} for different area overhead considerations. We can observe that even with 1\% area overhead, the network accuracy improves. However the improvements are significant only when the area overhead is greater than 8\% for both the correlation length values.

In summary, our results show that the proposed device-level optimization can help improve the tolerance of SPNNs of different scales under FPVs. Even for the cases where the nominal variation-free accuracy cannot be recovered using the optimized MZIs, our method can still be used to complement dynamic SPNN calibration methods to reduce their associated complexity, power consumption, and area overhead.

\section{Conclusion} \label{sec:: 6-Conclusion}
\begin{table}[t]
  \centering
  \caption{Network-2 accuracy with worst-case-tolerant MZIs designed using strip waveguides under width variations. }
  \begin{tabular}{|c|c|c|c|c|c|}
    \hline
    \rowcolor{Gray}
    Area & Correlation & Width & Arm & Pre-Opt & Post-Opt  \\
    \rowcolor{Gray}
    Overhead & Length & (nm) & Length &Accuracy &  Accuracy\\
    \rowcolor{Gray}
     & & & ($\mu$m) & & \\
    \hline
    \multirow{2}{*}{1\%} & 1~mm & \multirow{2}{*}{533} & \multirow{2}{*}{135.63} & 11.65\% & 20.47\% \\
    & 100~$\mu m$ &&& 54.27\% & 77.08\% \\
    \hline
    \multirow{2}{*}{2\%} & 1~mm & \multirow{2}{*}{589} & \multirow{2}{*}{136.19} & 11.65\% & 45.79\% \\
    & 100~$\mu m$ &&& 54.27\% & 86.48\% \\
    \hline
    \multirow{2}{*}{4\%} & 1~mm & \multirow{2}{*}{688} & \multirow{2}{*}{137.18} & 11.65\% & 83.97\% \\
    & 100~$\mu m$ &&& 54.27\% & 92.17\% \\
    \hline
    \multirow{2}{*}{8\%} & 1~mm & \multirow{2}{*}{853} & \multirow{2}{*}{138.83} & 11.65\% & 92.36\% \\
    & 100~$\mu m$ &&& 54.27\% & 93.77\% \\
    \hline
    \multirow{2}{*}{16\%} & 1~mm & \multirow{2}{*}{1111} & \multirow{2}{*}{141.41} & 11.65\% & 93.97\% \\
    & 100~$\mu m$ &&& 54.27\% & 94.28\% \\
    \hline
    \multirow{2}{*}{32\%} & 1~mm & \multirow{2}{*}{1200} & \multirow{2}{*}{142.3} & 11.65\% & 94.07\% \\
    & 100~$\mu m$ &&& 54.27\% & 94.29\% \\
    \hline
  \end{tabular}
  \label{Tab:: Phase Noise}
  \vspace{-1em}
\end{table}

In this paper, we have analyzed the impact of fabrication-process variations in the wavegide width and SOI thickness on coherent SPNNs. In particular, we have modeled undesired optical phase noises due to such variations at the MZI device level, and how such phase noises contribute to the performance degradation, for which we have considered relative variation distance ($RVD$) at the network level in optical unitary multipliers built using MZIs. Furthermore, we have proposed physical-level design optimization solutions to enhance MZI device tolerance under correlated FPVs in SPNNs. Our simulation results for two SPNN case studies of different sizes and considering realistic fabrication-process variation maps show that the proposed device-level optimization can help significantly improve SPNN inferencing accuracy. In addition, the results in this paper indicate the importance of considering variations during the design-phase of SPNNs to facilitate the application of online and dynamic calibration mechanisms in these networks, which are often power- and area-hungry.   

\bibliographystyle{IEEEtran}

\bibliography{IEEEabrv,References}

\end{document}